\documentstyle[preprint,revtex]{aps}
\begin {document}
\draft
\preprint{UCI-TR 93-6}
\begin{title}
Equations of Motion for Spinning Particles in External\\Electromagnetic
and Gravitational Fields
\end{title}
\author{Karl Yee and Myron Bander\cite{email1}}
\begin{instit}
Department of Physics, University of California, Irvine, California
92717, USA
\end{instit}

\receipt{February\ \ \ 1993}
\begin{abstract}
The equations of motion for the position and spin of a classical particle
coupled to an external electromagnetic and gravitational potential are
derived from an action principle. The constraints insuring a correct number
of independent spin components are automatically satisfied. In general
the spin is not Fermi-Walker transported nor does the position follow
a geodesic, although the deviations are small for most situations.
\end{abstract}
\pacs{PACS Numbers: 03.20.+i, 04.20.Fy}
\narrowtext
The derivation of the equations of motion of {\it classical} spinning
particles in external fixed fields has occupied physicists for over fifty
years. In special relativity it was first attacked by Frenkel
\cite{Frenkel}. Using his work as a basis Bargmann, Michel and Telegdi
\cite{BMT} discussed the precession of spinning particles in external
electromagnetic fields. It is amusing to note that, even today, there is
a controversy as to the torque and force on such particles in space and
time dependent fields \cite{Vaidman}. The discussion of a spinning particle
in an external gravitational potential goes back to Papapetrou
\cite{Papapetrou} who endowed a particle with spin by considering a
rotating mass-energy distribution in the limit of vanishing volume
but with the angular momentum remaining finite. Results were obtained  using
Grassmann variables and supersymmetry \cite{Ravndal} and an attempt at a
general procedure was made by Khriplovich \cite{Khriplovich}. Most of the
emphasis in the above works is on the equations for the spin components and,
with the exception of Refs.~\cite{Frenkel,Vaidman,Papapetrou}, the equations
for the motion of the position of the particle are either ignored or are
incomplete. We shall present a canonical procedure for obtaining the
equations of motion, both, for the spin and position. The results are
identical to those that would have been obtained using the method of
Ref.~\cite{Frenkel}.

We shall obtain the equations of motion in terms of the proper time,
$\tau$,  of the particle. The position of the particle will be denoted by
$x^{\mu}$ and the spin will be described by the antisymmetric spin matrix
$S_{ab}$. As usual, Greek indices will denote covariant vectors, tensors,
etc. and Latin ones those in local Lorentz frames; these are connected by
the vierbein field, $e_{\mu}^a(x)$. The spin matrix satisfies the Poisson
relation
\begin{equation}
\left\{ S_{ab},S_{cd}\right\}=\eta_{ac}S_{bd}+\eta_{bd}S_{ac}-
   \eta_{ad}S_{bc}-\eta_{bc}S_{ad}\, ,\label{poisson}
\end{equation}
with $\eta_{ab}$ the flat space metric. It will prove to be convenient to
obtain the equations of motion for the position from a Lagrangian and those
for the spin from a Hamiltonian; such a combined procedure calls for the
introduction of a Routhian \cite{classmech} ${\cal
R}(x^{\mu},S_{ab})$. At the end of this work we shall provide an
expression for the action. The equations of motion are
\begin{eqnarray}
\frac{\delta \int d\tau {\cal R}}{\delta x^{\mu}}&=&0\, ,\nonumber\\
\frac{dS_{ab}}{d\tau}&=&\left\{{\cal R},S_{ab}\right\}\, ;
  \label{eqmot}
\end{eqnarray}
$\tau$ is the proper time.
All the variables we have considered are not independent, but satisfy
various constraints. With $u^\mu$, the four velocity, these
are \begin{eqnarray}
u^{\mu}u_{\mu}&=&1\, ,  \label{velconst}\\
{S}_{ab}S^{ab}&=&2s^2\, ,\  \label{spinmagconst}\\
S_{ab}u^{a}&=&0\, . \label{spinconst}
\end{eqnarray}
Eq.~(\ref{velconst}) guarantees that
$\tau$ is the proper time and will be satisfied as long as ${\cal R}$ is
written in a reparametrization invariant form and
Eq.~(\ref{spinconst}) is satisfied. Eq.~(\ref{spinmagconst})
insures that the spin of the particle is constant and is equal to $s$;
it is automatically satisfied for all situations considered.
Eq.~(\ref{spinconst}) results from the fact that in the particle's rest
frame the spin tensor has only three independent components; it is this
constraint that causes all the complications.

For a particle in an external field derived from a vector potential
$A_{\mu}$ and a gravitational field specified by the spin connection
${\omega}_{\mu}^{ab}$ a tempting Routhian is
\widetext
\begin{equation}
{\cal R}_0=
m\sqrt{u^2}+eu^{\mu}A_{\mu}-\frac{u^{\mu}}{2}{\omega}_{\mu}^{ab}S_{a
b}
+\frac{eg}{4m}F^{ab}S_{ab}\sqrt{u^2}-\frac{\kappa}{8m}
R^{abcd}S_{ab}S_{cd}\sqrt{u^2}\, . \label{routhnaive} \end{equation}
\narrowtext
$m$ is the mass of the particle and $e$ is its charge; $g$ is the
gyromagnetic ratio and $\kappa$ specifies the strength of gravitational
magnetic moment coupling introduced in Ref.~\cite{Khriplovich}. This is the
most general Routhian not involving derivatives of the field strength
tensor or of the Riemann tensor and not involving terms of the form
$S_{ab}u^a$. Unfortunately, the equations of motion
derived using ${\cal R}_0$ do not satisfy the constraints of
Eq.~(\ref{spinconst}). There are two procedures that will guarantee this
constraint; both give the same equations of motion. One can follow the
method of Ref. \cite{Frenkel} and add to ${\cal R}_0$ the constraint
multiplied by a Lagrange multiplier. We shall follow a different procedure.
First define
\begin{equation}
{\tilde S_{ab}}=S_{ab}-S_{ac}\frac{u^cu_b}{u^2}-S_{cb}\frac{u^cu_a}{u^2}\, .
\end{equation}
The Poisson brackets of the ${\tilde S_{ab}}$'s is the one given in
Eq.~(\ref{poisson}) with the metric tensor $\eta_{ab}$ replaced by
$\eta_{ab}-u_au_b/u^2$. The ${\tilde S}_{ab}$ are related to the spin
vector $s^a$ by
\begin{eqnarray}
s^a&=&-\frac{1}{2}{\epsilon}^{abcd}{\tilde S}_{bc}u_d\, ,\nonumber\\
{\tilde S}_{ab}&=&{\epsilon}_{abcd}s^cu^d\, .\label{spinvector}
\end{eqnarray}
The spin vector satisfies $s^as_a=s^2$ and the constraint $s^au_a=0$.

A desired Routhian is obtained by replacing all $S_{ab}$'s
in Eq.~(\ref{routhnaive}) by ${\tilde
S_{ab}}$'s and by adding $\left ({du^a}/{d\tau}\right )S_{ab}u^b/u^2$
to it;
\begin{eqnarray}
{\cal R}&=&
m\sqrt{u^2}+eu^{\mu}A_{\mu}-\frac{u^{\mu}}{2}\omega_{\mu}^{ab}S_{ab}
+\frac{Du^a}{D\tau}\frac{S_{ab}u^b}{u^2}
\nonumber\\
&+&\frac{eg}{4m}F^{ab}{\tilde S}_{ab}\sqrt{u^2}-\frac{\kappa}{8m}
R^{abcd}{\tilde S}_{ab}{\tilde
S}_{cd}\sqrt{u^2}\, .
\label{routhian}
 \end{eqnarray}
Here, $D$ denotes a covatiant derivative and we have used the identity
\begin{equation}
\frac{du^a}{d\tau}\frac{S_{ab}u^b}{u^2}-\frac{u^{\mu}}{2}\omega_{\mu}^{ab}
{\tilde
S_{ab}}=\frac{Du^a}{D\tau}\frac{S_{ab}u^b}{u^2}-\frac{u^{\mu}}{2}
\omega_{\mu}^{ab}S_{ab}\, .
\end{equation}
The fourth term in Eq.~(\ref{routhian}) involves the
acceleration explicitly; it adds the Thomas precession term to the
equations of motion for the spin and insures that $S_{ab}u^a$ may be set
equal to zero.

The equations of motion, with Eqs.~(\ref{velconst}-\ref{spinconst})
satisfied, for the spin tensor are
\begin{eqnarray}
\frac{DS_{ab}}{D\tau}+\left (u_bS_{ac}-u_aS_{bc}\right )
\frac{Du^c}{D\tau}&=&-\left
(\frac{eg}{2m}F^{cd}+\frac{\kappa}{2m}R^{cdef}S_{ef}\right )
\nonumber\\ &\times &\left
[S_{ac}(\eta_{bd}-u_bu_d)+S_{bd}(\eta_{ac}-u_au_c)\right]\, .
\label{spineqmot}
\end{eqnarray}
This expression is consistent with Eq.~(\ref{spinmagconst}) and
Eq.~(\ref{spinconst}).
The equations of motion for the coordinates of the particle are
\begin{eqnarray}
&m\frac{Du_\mu}{D\tau}-eF_{\mu\nu}u^\nu-
\frac{1}{2}R_{\mu\nu}^{cd}S_{cd}u^\nu=
\frac{D}{D\tau}\left [\left (-\frac{eg}{4m}F^{cd}S_{cd}+
\frac{\kappa}{8m}R^{cdef}S_{cd}S_{ef}\right )u_\mu\right ]\nonumber\\
&\frac{D}{D\tau}\left (e_{\mu}^cS_{cd}\frac{Du^d}{D\tau}\right )
+\frac{eg}{4m}{F^{cd}}_{;\mu}S_{cd}-\frac{\kappa}{8m}{R^{cdef}}_{;\mu}
S_{cd}S_{ef}\nonumber\\
&+\frac{D}{D\tau}\left [e_{\mu}^h\left (\frac{eg}{2m}F^{cd}u_cS_{hd}-
\frac{\kappa}{2m}R^{cdef}S_{cd}u_eS_{hf}\right )\right ]\, .
\label{cmeqmot}
\end{eqnarray}
These equations are exact. Except in the case of large gravitational field
gradients the modifications due to the spin will be small \cite{MWT}. It
is interesting to study various limits of Eq.~(\ref{cmeqmot}). If we ignore
the right hand side of that equation and plug the results into
Eq.~(\ref{spineqmot}) we obtain
\begin{eqnarray}
\frac{DS_{ab}}{D\tau}=&-&\left
(\frac{eg}{2m}F^{cd}+\frac{\kappa}{2m}R^{cdef}S_{ef} \right )\left
(S_{ac}\eta_{bd}+S_{bd}\eta_{ac}\right )\nonumber\\ &+&\left
[\frac{e(g-2)}{2m}F^{cd}+\frac{(\kappa -1)}{2m}R^{cdef}S_{ef} \right ]\left
(S_{ac}u_bu_d+S_{bd}u_au_c\right )\, .\label{simpeqmot}
\end{eqnarray}
We know that the electromagnetic part of the equations of motion for the
spin simplify in the case $g=2$; we also see that there is a simplification
for the gravitational part in the case $\kappa=1$. That the Dirac equation
\begin{equation}
{\gamma}^a e_a^\mu\left (i\partial_\mu-
eA_\mu+\frac{1}{2}\omega_\mu^{cd}
S_{cd}\right )\psi -m\psi =0\, ,
\end{equation}
with $S_{cd}$ expressed in terms of the Dirac $\gamma$ matrices, yields
$g=2$ is well known; it also yields $\kappa=1$. We note that even in the
presence of only gravitational couplings, but with $\kappa\ne 0$, the
spin is {\it not} Fermi-Walker
transported \cite{MWT,Weinberg}. The corrections to Fermi-Walker transport
are very small, except as mentioned earlier, in the presence of large
gravitational fields and gradients.

Another interesting limit is the situation of no electromagnetic field and
$\kappa=0$. The equation of motion for the position of the particle is
\begin{equation}
m\frac{Du_\mu}{D\tau}-\frac{1}{2}R_{\mu\nu}^{cd}S_{cd}u^\nu-
\left (e_{\mu}^cS_{cd}\frac{D^2u^d}{D\tau^2}\right )=0\, .
\end{equation}
This agrees with the equations in Ref.~\cite{Papapetrou}; we note that
spinning particles do not follow geodesics. The term involving the time
derivative of the acceleration  has been interpreted, in
Ref.~\cite{Papapetrou}, as being responsible for classical {\it
Zitterbewegung\/}; in the following sense we agree with this interpretation:
in the absence of gravitational interactions a spinning particle will
oscillate in the plane perpendicular to the spin direction with a frequency
$\omega=E/s$; $s^2=S^{ab}S_{ab}/2$ is the magnitude of the spin vector. If we
set $|s|=\hbar /2$ we recover the quantum mechanical {\it Zitterbewegung}
frequency.

In the non-relativistic limit Eq.~(\ref{cmeqmot}), for a purely
magnetic dipole interaction is
\begin{equation}
m\frac{d\mbox{\boldmath $v$}}{dt}=\frac{eg}{2m}
\mbox{\boldmath $\nabla$}\left (\mbox{\boldmath
$H\cdot s$}\right ) +\frac{eg}{2m}\frac{d}{dt}\left (\mbox{\boldmath $E\times
s$} \right)\, ;
\end{equation}
${\bf s}$ is defined in Eq.~(\ref{spinvector}). This expression agrees
with the force equation advocated in Ref.~\cite{Vaidman}.

For completeness we present an action which corresponds to the
Routhian of Eq.~(\ref{routhian}). A convenient approach is to add a
Wess-Zumino \cite{WZ} term. For closed paths in proper time we
introduce a two dimensional manifold, ${\cal M}$, parametrized by $y_1,\ y_2$
whose boundary is the path $\tau$. The action is
\begin{equation}
{\cal A}=\frac{1}{s^2}\int_{\cal M}dy_1\
dy_2\epsilon^{\alpha\beta}\mbox{\rm Tr}
S\frac{\partial S}{\partial y_{\alpha}}\frac{\partial S}{\partial
y_{\beta}}+\int d\tau {\cal R}\, .
\end{equation}
The canonical equations obtained from the double integral yield the Poisson
brackets of Eq.~(\ref{poisson})

We wish to thank Professor A. Schwimmer for discussions on the
Wess-Zumino term. This work was supported in part by the National
Science Foundation under Grant No. PHY-9208386.

\end{document}